\documentclass[a4paper,fleqn,usenatbib]{mnras}

\usepackage{mathptmx}

\usepackage[T1]{fontenc}
\usepackage{ae,aecompl}

\usepackage{graphicx}	
\usepackage{amsmath}	
\usepackage{amssymb}	
\usepackage{times}
\usepackage{color}
\usepackage{float}
\usepackage{rotating}
\usepackage{url}

\def\apj{ApJ}
\def\mnras{MNRAS}
\def\aap{A\&A}
\def\apjl{ApJ}
\def\pasj{PASJ}
\def\nat{Nature}

\def\apjs{ApJS}

\def\aj{AJ}
\def\ssr{SSR}

\def\lesssim{\mathrel{\hbox{\rlap{\hbox{\lower4pt\hbox{$\sim$}}}\hbox{$<$}}}}
\let\la=\lesssim
\def\gtrsim{\mathrel{\hbox{\rlap{\hbox{\lower4pt\hbox{$\sim$}}}\hbox{$>$}}}}
\let\ga=\gtrsim

\title[Swift J1357.2-0933]{Swift J1357.2-0933: a massive black hole in the galactic thick disc}
\author[D. Mata S\'anchez et al.]{D. Mata S\'anchez $^{1,2}$\thanks{E-mail: dmata@iac.es}, T. Mu\~noz-Darias$^{1,2}$, J. Casares$^{1,2}$, J.~M. Corral-Santana $^{3}$ and \newauthor T. Shahbaz$^{1,2}$\\
$^{1}$Instituto de Astrof\'isica de Canarias, 38205 La Laguna, Tenerife, Spain\\
$^{2}$Departamento de astrof\'isica, Univ. de La Laguna, E-38206 La Laguna, Tenerife, Spain\\
$^{3}$ Instituto de Astrof\'isica, Pontificia Universidad Cat\'olica de Chile, Casilla 306, Santiago 22, Chile\\}

\date{Accepted 2015 September 8. Received 2015 September 7; in original form 2015 August 3 }

\pubyear{2015}

\begin{document}
\label{firstpage}
\pagerange{\pageref{firstpage}--\pageref{lastpage}}
\maketitle

\begin{abstract}
Swift J1357.2-0933 is one of the shortest orbital period black hole X-ray transients (BHTs). It exhibited deep optical dips together with an extremely broad H$\alpha$ line during outburst. We present 10.4-m GTC time-resolved spectroscopy during quiescence searching for donor star absorption features. The large contribution of the accretion flow to the total luminosity prevents the direct detection of the companion. Nevertheless, we constrain the non-stellar contribution to be larger than $\sim 80\%$ of the total optical light, which sets new lower limits to the distance ($d > 2.29\, \rm{kpc}$) and the height over the Galactic plane ($z>1.75\, \rm{kpc}$). This places the system in the galactic thick disc. We measure a modulation in the centroid of the H$\alpha$ line with a period of $P=0.11\pm0.04\, \rm{d}$ which, combined with the recently presented FWHM-$K_2$ correlation, results in a massive black hole ($M_1>9.3 \, \rm{M_\odot}$) and a $\sim$ M2V companion star ($M_2\sim 0.4\, \rm{M_\odot}$). We also present further evidence supporting a very high orbital inclination ($i\gtrsim 80^\circ$).

\end{abstract}



\section{Introduction}
\label{intro}

Galactic X-ray binaries offer the best opportunity to detect stellar mass black holes (BHs) through the accretion luminosity produced by material from the companion star falling into its gravitational well. The vast majority of BHs are detected in transient low-mass X-ray binaries (LMXBs), which spend most part of their lives in a faint, quiescent state (about $\rm 10^{30-34}\, erg\, s^{-1}$; see e.g. \citealt{Armas-Padilla2014b}). They are discovered during occasional outburst, where their X-ray luminosity increases above $\sim 10$ per cent of the Eddington limit. There are only 17 dynamically confirmed BHTs, as well as $\sim 33$ candidates (see \citealt*{Casares2014}; \citealt{Corral-Santana2015}).

Swift J1357.2-0933 (hereafter J1357) is a LMXB X-ray transient with galactic coordinates $l=328.702 {\,}^{\circ}$ and $b=+50.004{\,}^{\circ}$ \citep{Krimm2011}, whose peak X-ray luminosity place it in the very faint regime (see \citealt{Armas-Padilla2013} for an X-ray analysis during outburst). The detection of optical dips both in outburst (\citealt{Corral-Santana2013}, hereafter CS13) and quiescence \citep{Shahbaz2013} suggests J1357 is a very high inclination system. The orbital period is among the shortest of its class ($P= 2.8\, \rm{h}$, CS13) and the mass function is constrained to be $f(M_1)> 3.0\, \rm{M_\odot}$, where the radial velocity of the donor ($K_2$) was estimated from the double-peak separation of the H$\alpha$ profile \citep{Orosz1994, Orosz1995}. This strongly advocates for the presence of a BH. \citet*{Rau2011} proposed a tentative distance to the system of $d\sim 1.5\,\rm{kpc}$ considering the donor star to be the dominant source of quiescent light in the optical and near-infrared regime. However, the quiescent spectral energy distribution (SED) is best described by a single power-law model, which suggests little, if any, thermal contribution from the secondary star \citep{Shahbaz2013}. This implies that only lower limits to the distance can be obtained unless the donor star contribution is properly characterized.


\section{Observations}
\label{observations}

\begin{figure*}
\includegraphics[width=168mm]{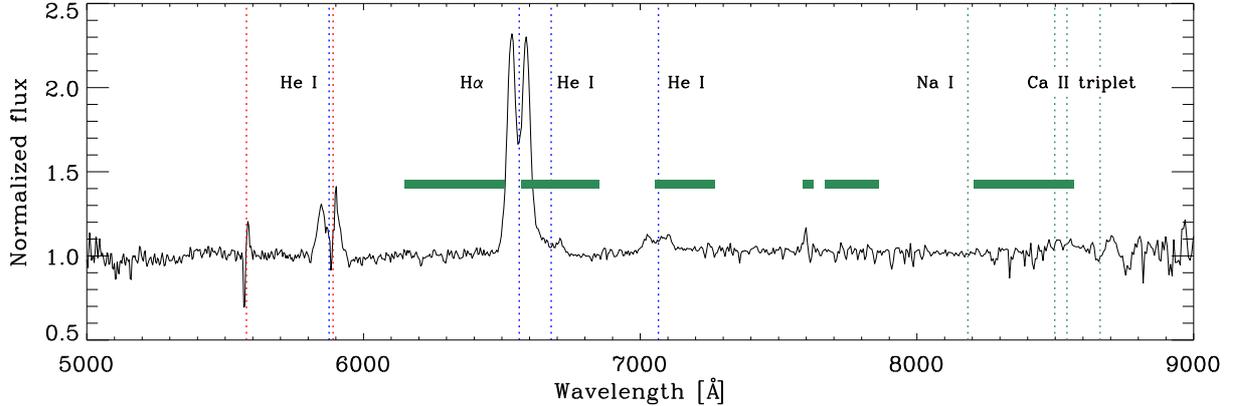}
\caption{Averaged spectrum of J1357. Accretion disc emission features (H$\alpha$ and \ion{He}{I}) are marked with blue, dotted vertical lines. Regions contaminated by sky subtraction residuals are plotted as red, doted vertical lines. The expected M-type donor star absorption features are shown as green bands (TiO molecular bands) and green, dotted vertical lines (\ion{Na}{I} and \ion{Ca}{II} triplet).
}
\label{figaverage}
\end{figure*}

J1357 was observed with the Optical System for Imaging and low-Intermediate-Resolution Integrated Spectroscopy (OSIRIS) located in the Nasmyth-B focus of the 10.4-m GTC, in La Palma (Spain). We used the R500R optical grism combined with a 1.0'' slit ($\rm R=352,\, 4.88$ {\AA}/pix), covering the spectral range $\rm 4800-10000$ {\AA}. Observations were obtained in four different nights within April to June 2014.

Eleven consecutive spectra per night were acquired at airmasses in the range $\sim 1.3 - 1.7$, with individual exposure times of $\rm 875\, s$ ($\rm \sim 3\, h$ per night) sampling the proposed $\rm 2.8\, h$ orbital period. We used IRAF\footnote{IRAF is distributed by National Optical Astronomy Observatories, operated by the Association of Universities for Research in Astronomy, Inc., under contract with the National Science Foundation.} standard routines for bias and flat-field corrections. The slit was rotated to $\rm{PA}=-53.16^\circ$ in order to allow for simultaneous observations of the object and a nearby early-type dwarf star located at $2.06\, \rm{arcmin}$ NE of the target. The wavelength calibration was obtained from observations of calibration arcs and instrumental flexure was calculated from the drift of the [\ion{O}{I}] $\rm 5577$ {\AA} sky line and subsequently corrected in every individual spectrum.

The spectrum of the calibration star is classified as a F8V after comparison with templates from the \textit{MILES} spectral library (\citealt{Sanchez-Blazquez2006}, \citealt{Falcon-Barroso2011}) using $\chi^2$ minimization routines within MOLLY (e.g. \citealt{Casares1996}). In order to maximize the chances of detecting donor star absorption features in J1357, we produced a set of telluric spectra by removing the F8V spectral lines, which were subsequently substracted from the target to produce spectra free of telluric features (e.g. \citealt{Beekman2000}).

The above strategy yielded four sets of eleven spectra. Points deviating more than $\rm 3\sigma$ above or below the continuum, probably caused by bad pixels or residuals from the sky subtraction, were interpolated. One spectrum was discarded on each of the two first nights due to very poor signal-to-noise ratio (S/N) caused by the presence of clouds. Absorption features from the donor star are not evident in any of the spectra, nor in the combined spectrum (see Fig. \ref{figaverage}).

\section{Results}
\label{results}

Using the 42 GTC spectra we have determined the veiling factor, the orbital period and the main properties of the H$\alpha$ line in quiescence.

\subsection{Search for companion star features: Skew-mapping}

The skew-mapping technique has been applied in cataclysmic variables to detect weak companion star features (e.g. \citealt{Vande2003}, \citealt{Smith2005}).
Our analysis compares a template M3V star spectrum with different Doppler-corrected averaged spectra created from our database. Each averaged spectrum is produced by considering a specific zero phase in the range 0-1 (we inspect all values with a 0.05 step) and velocity of the companion star between $K_2= 0-2000\, \rm{km\, s^{-1}}$, in $100\, \rm{km\, s^{-1}}$ steps. Individual spectra are shifted and coadded assuming the orbital period of CS13. The cross-correlation with the M3 template star is expected to reveal a significant peak above noise at the correct parameters for the companion star motion. This technique did not reveal any preferred value in the \textit{$K_2$ - zero phase} parameter space, not even after considering several different periods around the one proposed in CS13 ($0.09-0.15\, \rm{d^{-1}}$, steps of $0.001\, \rm{d^{-1}}$). Therefore, we conclude that no companion star features are present in our spectra.

\subsection{Veiling factor}
\label{veiling}
Magnitudes in the OSIRIS $r'$ filter were obtained from the acquisition images of each night: $r'_1=21.71\pm  0.08;\quad r'_2=21.45\pm  0.06;\quad r'_3=20.84\pm  0.08 ;\quad r'_4=21.00\pm  0.11$. The first two values (April 29th and 30th) are similar to those reported in \citet[][$r'=21.54\pm 0.35$]{Shahbaz2013}. On the other hand, the latest two nights (June 2nd and 28th) reveal a somehow brighter system (but still in quiescence). We note that although this might be caused by the known strong short-term variability displayed by the system \citep{Shahbaz2013}, the spectra of the first two nights are clearly noisier under similar sky conditions, reflecting a true drop in brightness.

The absence of companion star features imposes a minimum constraint to the veiling factor ($X$), defined as the fractional contribution of the accretion related luminosity ($L^{r'}_{\rm{acc}}$) to the total flux ($L^{r'}_{\rm{acc}}+L^{r'}_2$, where $L^{r'}_2$ is the donor star luminosity) in the OSIRIS $r'$ filter wavelength range ($\rm 5500-7400$ {\AA}). On the other hand, by considering the relation between the orbital period and the mean density presented in \citet*{Faulkner1972}, and tabulated values for main-sequence stars \citep{Cox2000}, the spectral type of the companion star is constrained to be M2V or later. We compared our spectrum with templates of main-sequence stars from the \textit{MILES} Spectral Library. For each spectral type, we measured the normalized flux of the deepest photospheric absorption line present in the template spectrum. The veiling necessary to make these features shallower than the noise level ($3\sigma$) within the corresponding spectral region of the J1357 spectrum is a lower limit to the veiling factor (see \citealt{MataSanchez2015} for a similar analysis of NIR spectra). 

We find that the required veiling factor for a $\rm M2V$ star to be swamped by the accretion flux is $X>0.81$. This limit is also valid if the donor is evolved and its spectral type is later than $\rm M2V$ (we have explored spectral types as late as $\rm M6V$ ). This result is obtained using the averaged quiescent spectrum of the first two nights where the system is fainter, and therefore, a higher contribution of the companion star is expected.

We have explored spectral types as late as $\rm M6V$, obtaining similar results than for $\rm M2V$. We note that using slightly different tabulated values for dwarf star mean densities (e.g. \citealt{Carroll2007}), does not affect our results.

\subsection{Orbital period from a two-Gaussian model fit}
\label{period}

We fit the H$\alpha$ double-peaked emission profile with a model consisting of two Gaussians with equal FWHM. The height of the Gaussians and offset with respect to the H$\alpha$ rest wavelength ($\rm 6562.78$ {\AA}) are left as free parameters.

 \begin{figure}
\includegraphics[width=85mm]{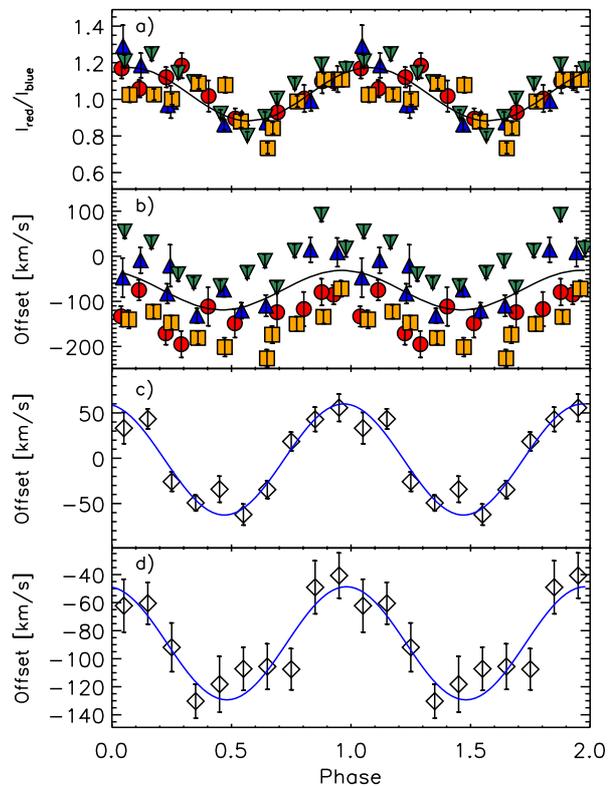}
\caption{Orbital evolution of several H$\alpha$ parameters obtained from our two-Gaussian model fit (three top panels) and from a diagnostic diagram (bottom panel) folded with the best fit period obtained from periodogram analysis. Panel a: intensity ratio between red and blue peak. Best fit is represented as a black, solid line. Panel b: radial velocities of the non-detrended centroid as measured by the two-Gaussian model. In both panels, each night is represented by different symbols and colours as follows: red circles (April 29th), blue triangles (April 30th), green upside-down triangles (June 2nd) and orange squares (June 28th). The best sinusoidal fit to the non-detrended data is also shown. Panel c: radial velocities of the line centroid obtained from $10$ phase bin averaged spectra after de-trending. The best sinusoidal fit is shown as a blue, solid line. Panel d: same as previous panel but using radial velocity curves from the diagnostic diagram (Gaussian separation of $a=4500\, \rm{km\,s^{-1}}$). Data are plotted over two phase cycles for clarity.
}
\label{figperfold}
\end{figure}

\subsubsection*{H$\alpha$ centroid velocity}

The mean of the Gaussians offsets is expected to trace the motion of the accretion disc around the centre of mass of the system. We note that the mean offset value for each night was significantly different (see Fig. \ref{figperfold}b). This behaviour has been reported in other BHTs, for example in XTE J1118+480 it is explained as a result of disc precession \citep{Torres2002}. Recently, \citet{Torres2015} detected variations in the systemic velocity of J1357. They did not consider this result to support the precessing disc scenario, but interpreted as systematic effects. However, our longer database suggests that instead this is a consequence of a physical process taking place in the system. In order to fit the whole sample, we then decided to de-trend the mean nightly offsets. 

\begin{figure}
\includegraphics[width=85mm]{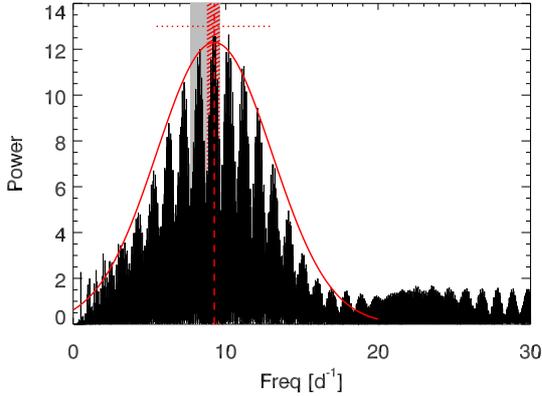}
\caption{Periodogram obtained with the Lomb-Scargle method. Red, solid line shows a Gaussian fit to the highest group of peaks. The red, dotted horizontal line represents the $1\sigma$ uncertainty ($P=0.11\pm 0.04\,\rm{d}$). We inspected frequencies up to $60\, \rm{d^{-1}}$, but the power spectrum for frequencies over $30\, \rm{d^{-1}}$ is nearly flat. The red, dashed vertical line refers to the strongest peak consistent with the outburst period proposed by CS13. Grey band depicts the constraint to the period ($P=0.117\pm 0.013\,\rm{d}$) from outburst data (CS13). The red, stripped band refers to the standard deviation obtained from a Gaussian fit to the family of peaks inside the grey band ($P=0.109\pm 0.005\,\rm{d}$).
}
\label{figpgram}
\end{figure}

We performed a Lomb-Scargle normalised periodogram of the (de-trended) mean offset variations and the result is presented in Fig. \ref{figpgram}. We used the \textit{period} time-series analysis package \citep*{Dhillon2001}. A Gaussian fit to the broad peak with the highest power yields a period of $P=0.11\pm   0.04\, \rm{d}$ ($1\, \sigma$). This is consistent with the one measured by CS13 using a longer database with higher S/N during outburst ($P=0.117\pm 0.013\,\rm{d}$). We find two families of peaks consistent with the outburst period. Gaussian fits result in $P=0.109\pm 0.005\,\rm{d}$ and $P=0.121\pm 0.006\,\rm{d}$, respectively. The former is favoured by its highest amplitude (depicted as a red, stripped band in Fig. \ref{figpgram}). Nevertheless, further observations are necessary to confirm this result. We note that comparable results are obtained by performing sinusoidal fits and minimise chi-square.

The data were folded on the previously obtained period and are shown in Fig. \ref{figperfold}b, where different colours and symbols are used to mark different systemic velocities observed on each night. We subsequently performed a non-linear least squares fit to the data using the following sinusoidal function:
$$V=\gamma + K_1 \sin{(2\pi(\phi-\phi_{~0}))}$$
We obtained $K_1=44\pm 3 \, \rm{km\, s^{-1}}$, $\gamma=75\pm 2 \, \rm{km\, s^{-1}}$ and $\phi_{~0} = 0.71\pm 0.01$. Even if it might be qualitatively acceptable, the high reduced chi-square value ($\chi_\mathrm{~r}^2$=17.26) shows that the non de-trended data does not properly fit to a sinusoid, and de-trending is necessary for an optimal analysis. Subsequently, the original spectra were combined into $10$ phase bins after removing each night's mean value. The fit results in $\phi_{~0} = 0.720\pm 0.009$ and $K_1=61\pm 4 \, \rm{km\, s^{-1}}$ for $\chi_\mathrm{~r}^2$=1.30 (see Fig. \ref{figperfold}c).

\subsubsection*{Intensity ratio and double peak separation}

The intensity ratio of the peaks of the H$\alpha$ profile (defined as $I_{\rm{red}}/I_{\rm{blue}}$) exhibits a periodicity similar to that observed in the centroid offsets. Folding these data onto the previously obtained period and fitting a sinusoid results in $\chi_\mathrm{~r}^2$=5.12 (see Fig. \ref{figperfold}a). \citet{Torres2015} reported asymmetric H$\alpha$ profiles, but no recurrent variability was observed. The detection of this periodic evolution of the profile along four different nights suggest an scenario with an asymmetric outer disc structure, perhaps a hot-spot on a tidal arm. This would also explain the deviation observed from a perfect fit, since other processes affecting the intensity ratio might be at play.

The double-peak separation does not show significant variability, and it is consistent with a constant value. The weighted average of the double-peak separation is $D_p=2430\pm   50 \, \rm{km\, s^{-1}}$, significantly larger than the one measured in outburst by CS13 ($D_p=1790\pm   67 \, \rm{km\, s^{-1}}$). This implies that the outer disc velocity ($v_d=1215\pm 25\, \rm{km\, s^{-1}} $) is higher in quiescence, which is an expected behaviour since the disc expands during outburst, reaching areas with lower velocities. The reported value is consistent within $2\sigma$ with that reported by \citet[][$D_p=2340\pm   20 \, \rm{km\, s^{-1}}$]{Torres2015}.  

\subsection{The diagnostic diagram}
\label{dgmethod}

The broad wings of the H$\alpha$ profile are supposed to trace the motion of the compact object. We apply the diagnostic diagram (aka double Gaussian technique \citealt*{Shafter1986}), aiming at constraining radial velocity variations of the H$\alpha$ wings. We inspected several Gaussian separations ($ a=2500-5000\,\rm{km\, s^{-1}}$) to perform a diagnostic diagram analogous to that presented in CS13. Unfortunately, our limited database and the lower S/N ratio prevents us from obtaining conclusive determinations for the $K_1$ and $\gamma$ values. A wide plateau on the evolution of the fitted parameters from $ a=4000\,\rm{km\, s^{-1}}$, combined with the inspection of the individual radial velocity curves, suggests $ a=4500\,\rm{km\, s^{-1}}$ as the preferred value (see the folded, radial velocity curve for this separation in Fig. \ref{figperfold}d). 

Mean values of the compact object orbital velocity ($K_1=40\pm 12 \,\rm{km\, s^{-1}}$) and systematic velocity of the system ($\gamma = -79 \pm 13 \,\rm{km\, s^{-1}}$) are obtained from sinusoidal fits to radial velocity curves in the range $ a=4000-5000\,\rm{km\, s^{-1}}$ (uncertainties refer to standard deviations). This result is consistent with previous determinations of the systemic velocity in quiescence ($\gamma = -130 \pm 50 \,\rm{km\, s^{-1}}$; \citealt{Torres2015}) and the compact object's velocity reported in outburst ($K_1=43\pm 2 \,\rm{km\, s^{-1}}$; CS13). However, the systemic velocity value obtained from the outburst periodogram ($\gamma \sim -150 \,\rm{km\, s^{-1}}$, CS13) does not seem to be in agreement with our results.  

We note that, in contrast to the two-Gaussian profile modelling (see Section \ref{period}), nigthly de-trending is not necessary for the radial velocity curves obtained with the double Gaussian technique. This can be explained by considering that the two-Gaussian modelling traces the core of the H$\alpha$ double peak profile, and therefore, outer parts of the disc where disc preccesion effects might be important. However, the diagnostic diagram is sensitive to the emission line wings, which traces inner parts of the disc less affected by companion star tidal forces. 

\subsection{H-alpha full-width-at-half-maximum}
We have measured the full-width-at-half-maximum (FWHM) of the H$\alpha$ emission profile in each individual spectrum. After substracting quadratically the instrumental resolution, the average value is $\rm{FWHM}=4152\pm 209 \, \rm{km\,s^{-1}}$, where the uncertainty reflects the standard deviation of the 42 measures. This value is larger than the one measured in outburst ($\rm{FWHM}\sim 3300 \, \rm{km\,s^{-1}}$, CS13) and consistent with the quiescence value reported in \cite{Torres2015} and \citet{Casares2015}, i.e. $\rm{FWHM} = 4025 \pm 110 \, \rm{km\,s^{-1}}$ and $\rm{FWHM} = 4085 \pm 328 \, \rm{km\,s^{-1}}$ respectively. The previously reported broadest FWHM corresponds to XTE J1118+480 ($\rm{FWHM}\sim 2500 \, \rm{km\,s^{-1}}$, \citealt{Torres2004}). 

\section{Discussion}
\label{discussion}
Observations taken during the outburst decay have sometimes resulted in wrong orbital period determinations (e.g. V404 Cygni; \citealt*{Casares1992}). However, in our case, the orbital period measured from quiescent spectra is fully consistent with the outburst results, supporting CS13 conclusions.
\citet{Casares2015} has recently presented a correlation between the FWHM of H$\alpha$ and the orbital velocity of the companion star $K_2 \simeq 0.233(13)\, \rm{FWHM}$. For the case of J1357 this results in a donor star velocity of $K_2 = 967\pm 49 \, \rm{km\,s^{-1}}$. Note that this value is consistent with the independent empirical relation derived for quiescent BHT between the outer disc velocity ($v_d$) and $K_2$ \citep{Orosz1994,Orosz1995}, which results in $K_2>806\, \rm{km\,s^{-1}}$ (see also \citealt{Torres2015}). This, combined with CS13 orbital period (fully consistent with our results) yields $f(M_1)=11.0 \pm 2.1 \, \rm{M_\odot}$ (i.e. $M_{BH}>8.3\, \rm{M_\odot}$ at 90\% confidence). If we include the BH orbital velocity estimated in CS13 ($K_1=43\pm 2 \, \rm{km\,s^{-1}}$), the obtained mass ratio ($q=M_2/M_1 \sim 0.04$) provides more restrictive limits to both $M_{BH}$ and $M_2$. We find $M_{BH}\geq 9.3 \, \rm{M_\odot}$ and $M_2\geq  0.4\, \rm{M_\odot}$. Here, we use the limit value for the orbital inclination $i=90^\circ$. This conservative result places the system as the most massive LMXB BH together with GRS 1915+105 ($\rm 10.1\pm 0.6\, M_\odot$, \citealt{Steeghs2013}). Only Cyg X-1, a high-mass X-ray binary, exceeds the BH mass presented here ($\sim 15 \, \rm{M_\odot}$, \citealt{Orosz2011}). The donor star mass is consistent with the constraint on the spectral type (later than M2V; see Sec. \ref{veiling}), which requires $M_2 \lesssim 0.4 \, \rm{M_\odot}$.

\subsection{Distance and height over the Galactic plane}
The constraint on the veiling factor, combined with the spectral type (mass) of the donor star, imposes a more restrictive lower limit on the distance to the system. Photometric values were obtained from acquisition images (one per night) in the $r'$-band. We use the relation between $r'$, $B$ and $V$ magnitudes described in \citet{Fukugita1996}, combined with $M_V$ and $B-V$ tabulated values of main-sequence stars \citep{Cox2000} to obtain absolute magnitudes, $M_{r'}$. If we consider the latest spectral type for the donor star proposed by CS13 ($\rm {M5V}$, $M_{r'}=11.61$), we obtain a conservative lower limit to the distance of the system. This assumes a somewhat evolved donor from the \citet{SmithDhillon1998} empirical relation, instead of a main sequence star. We measure $r'=21.6\pm 0.2$ from the two nights when the system is faintest, which combined with $X>0.81$ results in a conservative constraint to the distance of $d > 2.29\, \rm{kpc}$. If we consider the companion star spectral type obtained in the previous section ($\rm {M2V}$, $ M_{r'}=9.28$), the system is required to be further than $d > 6.7\, \rm{kpc}$, a value which is close to the upper limit reported in \citet{Shahbaz2013}.

The above constraint to the distance sets new lower limits to the X-ray luminosity during both outburst and quiescence. \citet{Armas-Padilla2014b, Armas-Padilla2013} obtained $L_{\rm{quies}}=8\times 10^{29}\, \rm{erg\, s^{-1}}$ and $L_{\rm{peak}}=10^{35}\, \rm{erg\, s^{-1}}$ considering a distance of $d=1.5\,\rm{kpc}$. Our constraint implies: $L_{\rm{quies}}>1.9\times 10^{30}\, \rm{erg\, s^{-1}}$ and $L_{\rm{peak}}>2.33\times 10^{35}\, \rm{erg\, s^{-1}}$. These values are still consistent with J1357 being a very faint X-ray transient and perhaps the faintest stellar mass BH known in quiescence (see \citealt{Armas-Padilla2014b} for a discussion on the topic).

The lower limit to the distance, combined with the high Galactic latitude ($b=50.004^{\rm{\circ}}$), places the system at $z>1.75\, \rm{kpc}$ above the Galactic plane. Other BHT that might be members of the Galactic thick disc population ($z \gtrsim 1\, \rm{kpc}$, \citealt*{Gilmore1983}) are: SWIFT J1357.2-0933 (this work and CS13), MAXI~J1659-152 (BH candidate, \citealt{Kuulkers2013}), XTE~J1118+480 \citep{Uemura2000,Gelino2006}, XTE~J1859+226 \citep{Corral-Santana2011}, H1705-250 (\citealt{Remillard1996}, \citealt*{Jonker2004}), GS~1354-64 \citep{Casares2004,Casares2009} and SWIFT J1753.5-0127 (BH candidate, \citealt{Zurita2008}); see \citet{Corral-Santana2015}.


\subsection{On the orbital inclination}

\begin{figure}
\includegraphics[width=85mm]{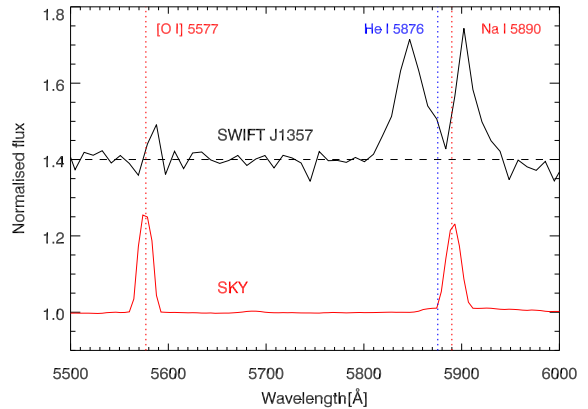}
\caption{Spectra centred on the \ion{He}{i} ($\rm 5876$ {\AA}) emission line. Upper spectrum: normalised, averaged J1357 spectrum of the two brightest nights after sky correction. Lower spectrum: Averaged spectrum of the sky multiplied by a constant factor in order to compare with the upper spectrum.
}
\label{figeclipse}
\end{figure}

The \ion{He}{i} $\rm 5876$ {\AA} emission line exhibits a particular profile with a sharp, deep absorption core, even reaching $0.95$ the continuum value in some spectra. This has to be treated with caution since the \ion{Na}{i} $\rm 5890$ {\AA} sky line is placed close to the core of the double-peaked \ion{He}{i} profile, and therefore a deficient sky subtraction could affect the profile. In order to minimize this effect we combined the data of the two nights were the object was brighter, because this allows for a better correction. The final extracted spectrum (see Fig. \ref{figeclipse}) still exhibits a sharp core profile, reaching $\sim 1.05$ times the continuum level. The residual sky substraction of the nearest and equally intense sky line [\ion{O}{i}] $\rm 5577$ {\AA} is reduced to almost noise level. This result suggests that the deep absorption core is a real feature.

Deep \ion{He}{i} line absorption cores have only been previously observed in eclipsing, high inclination ($i\gtrsim 75^\circ$, \citealt*{Schoembs1983}) cataclysmic variables in quiescence e.g. Z Cha, \citep*{Marsh1987}, which has $i=81^\circ$. 
Obscuring material above the plane of the accretion disc has been proposed as the origin of the deep, central absorption observed in He I and Balmer hydrogen lines (see \citealt*{Rayne1981}). Therefore, it should not be surprising to find similar behaviour in high inclination BHTs. However, only systems with inclinations up to $i \sim 70^\circ$ have been discovered and detected in quiescence so far (e.g. XTE J1118+480, \citealt{Gelino2006}). None of them shows such deep absorption cores. \citet{Torres2015} already noticed depth variations in the core of of H$\alpha$ profile. We also observe variations in the normalized flux of the H$\alpha$ core in the range 1.3 to 2.1 times the continuum value.

It has been suggested by \citet{Armas-Padilla2014a} and \citet{Torres2015} that the X-ray properties of the source (e.g. absence of emission/absorption lines) and the low extinction, comparable to that expected from interstellar material in the line-of-sight, argue against the very high inclination ($i \gtrsim 80^\circ$) scenario proposed by CS13. Given the lack of eclipsing BHs, the closest comparison can be made with high inclination neutron star systems. These only show absorption lines during thermal, soft states (\citealt*{Ponti2014}; \citealt*{Ponti2015}). However, it should be noted that J1357 never abandoned the hard state (see \citealt{Munoz-Darias2014} for a direct comparison between neutron star and BH states). On the other hand, the two eclipsing accretion disc corona sources X1822-371 and 2S 0921-63, show extinction values consistent with an interstellar origin (i.e no intrinsic extinction) when fitted with standard hard state spectral models (\citealt{Iaria2001}; \citealt{Kallman2003}). This suggests that X-ray photons emitted in the direction of the outer disc rim do not reach the observer, and only those radiated above/below the rim do it, solely interacting with the interstellar material in the line-of-sight. Therefore, we still consider $i \geq 80^\circ$ as the most plausible value. The detection of \ion{He}{i} line cores and the already large BH mass (for $i=90^\circ$) presented here further support this conclusion.  

\section{Conclusions}
\label{conclusion}

We have used 10.4-m GTC optical spectra of Swift J1357.2-0933 to constrain the accretion related contribution to the optical emission in quiescence. This yields more restrictive lower limits to the distance ($d > 2.29\, \rm{kpc}$) and height over the Galactic plane ($z>1.75\, \rm{kpc}$), placing the system in the Galactic thick disc. We detect variability in the H$\alpha$ profile modulated with a period $P=0.11\pm0.04\, \rm{d}$, confirming the period detected in outburst. Using the recently presented $\rm FWHM-K_2$ correlation, new constraints to the fundamental parameters are derived. In particular, we obtain $M_{BH}>9.3 \, \rm{M_\odot}$. We also favour $q\sim 0.04$ and $M_2\sim 0.4\, \rm{M_\odot}$. This indicates that Swift J1357.2-0933 harbours one of the most massive BHs known in our Galaxy.

\section*{Acknowledgements}
\label{acknowledgements}
We are thankful to Gabriele Ponti for useful discussion on the X-ray properties of the source. DMS acknowledges Fundaci\'on La Caixa for the financial support received in the form of a PhD contract. JMC-S acknowledge financial support to CONICYT through the FONDECYT project No. 3140310. We also acknowledge support by the Spanish Ministerio de Econom\'ia y competitividad (MINECO) under grant AYA2013-42627. MOLLY software developed by T. R. Marsh is gratefully acknowledged. 

\bibliographystyle{mnras} 
\bibliography{MiBiblio.bib}

\begin{thebibliography}{}
\makeatletter
\relax
\def\mn@urlcharsother{\let\do\@makeother \do\$\do\&\do\#\do\^\do\_\do\%\do\~}
\def\mn@doi{\begingroup\mn@urlcharsother \@ifnextchar [ {\mn@doi@}
  {\mn@doi@[]}}
\def\mn@doi@[#1]#2{\def\@tempa{#1}\ifx\@tempa\@empty \href
  {http://dx.doi.org/#2} {doi:#2}\else \href {http://dx.doi.org/#2} {#1}\fi
  \endgroup}
\def\mn@eprint#1#2{\mn@eprint@#1:#2::\@nil}
\def\mn@eprint@arXiv#1{\href {http://arxiv.org/abs/#1} {{\tt arXiv:#1}}}
\def\mn@eprint@dblp#1{\href {http://dblp.uni-trier.de/rec/bibtex/#1.xml}
  {dblp:#1}}
\def\mn@eprint@#1:#2:#3:#4\@nil{\def\@tempa {#1}\def\@tempb {#2}\def\@tempc
  {#3}\ifx \@tempc \@empty \let \@tempc \@tempb \let \@tempb \@tempa \fi \ifx
  \@tempb \@empty \def\@tempb {arXiv}\fi \@ifundefined
  {mn@eprint@\@tempb}{\@tempb:\@tempc}{\expandafter \expandafter \csname
  mn@eprint@\@tempb\endcsname \expandafter{\@tempc}}}

\bibitem[\protect\citeauthoryear{{Armas Padilla}, {Degenaar}, {Russell}  \&
  {Wijnands}}{{Armas Padilla} et~al.}{2013}]{Armas-Padilla2013}
{Armas Padilla} M.,  {Degenaar} N.,  {Russell} D.~M.,   {Wijnands} R.,  2013,
  \mn@doi [\mnras] {10.1093/mnras/sts255}, \href
  {http://adsabs.harvard.edu/abs/2013MNRAS.428.3083A} {428, 3083}

\bibitem[\protect\citeauthoryear{{Armas Padilla}, {Wijnands}, {Altamirano},
  {M{\'e}ndez}, {Miller}  \& {Degenaar}}{{Armas Padilla}
  et~al.}{2014a}]{Armas-Padilla2014a}
{Armas Padilla} M.,  {Wijnands} R.,  {Altamirano} D.,  {M{\'e}ndez} M.,
  {Miller} J.~M.,   {Degenaar} N.,  2014a, \mn@doi [\mnras]
  {10.1093/mnras/stu243}, \href
  {http://adsabs.harvard.edu/abs/2014MNRAS.439.3908A} {439, 3908}

\bibitem[\protect\citeauthoryear{{Armas Padilla}, {Wijnands}, {Degenaar},
  {Mu{\~n}oz-Darias}, {Casares}  \& {Fender}}{{Armas Padilla}
  et~al.}{2014b}]{Armas-Padilla2014b}
{Armas Padilla} M.,  {Wijnands} R.,  {Degenaar} N.,  {Mu{\~n}oz-Darias} T.,
  {Casares} J.,   {Fender} R.~P.,  2014b, \mn@doi [\mnras]
  {10.1093/mnras/stu1487}, \href
  {http://adsabs.harvard.edu/abs/2014MNRAS.444..902A} {444, 902}

\bibitem[\protect\citeauthoryear{{Beekman}, {Somers}, {Naylor}  \&
  {Hellier}}{{Beekman} et~al.}{2000}]{Beekman2000}
{Beekman} G.,  {Somers} M.,  {Naylor} T.,   {Hellier} C.,  2000, \mn@doi
  [\mnras] {10.1046/j.1365-8711.2000.03572.x}, \href
  {http://adsabs.harvard.edu/abs/2000MNRAS.318....9B} {318, 9}

\bibitem[\protect\citeauthoryear{{Carroll} \& {Ostlie}}{{Carroll} \&
  {Ostlie}}{2006}]{Carroll2007}
{Carroll} B.~W.,  {Ostlie} D.~A.,  2006, {An introduction to modern
  astrophysics and cosmology}

\bibitem[\protect\citeauthoryear{{Casares}}{{Casares}}{2015}]{Casares2015}
{Casares} J.,  2015, \apj, 808, 80

\bibitem[\protect\citeauthoryear{{Casares} \& {Jonker}}{{Casares} \&
  {Jonker}}{2014}]{Casares2014}
{Casares} J.,  {Jonker} P.~G.,  2014, \mn@doi [\ssr]
  {10.1007/s11214-013-0030-6}, \href
  {http://adsabs.harvard.edu/abs/2014SSRv..183..223C} {183, 223}

\bibitem[\protect\citeauthoryear{{Casares}, {Charles}  \& {Naylor}}{{Casares}
  et~al.}{1992}]{Casares1992}
{Casares} J.,  {Charles} P.~A.,   {Naylor} T.,  1992, \mn@doi [\nat]
  {10.1038/355614a0}, \href {http://adsabs.harvard.edu/abs/1992Natur.355..614C}
  {355, 614}

\bibitem[\protect\citeauthoryear{{Casares}, {Mouchet}, {Martinez-Pais}  \&
  {Harlaftis}}{{Casares} et~al.}{1996}]{Casares1996}
{Casares} J.,  {Mouchet} M.,  {Martinez-Pais} I.~G.,   {Harlaftis} E.~T.,
  1996, \mnras, \href {http://adsabs.harvard.edu/abs/1996MNRAS.282..182C} {282,
  182}

\bibitem[\protect\citeauthoryear{{Casares}, {Zurita}, {Shahbaz}, {Charles}  \&
  {Fender}}{{Casares} et~al.}{2004}]{Casares2004}
{Casares} J.,  {Zurita} C.,  {Shahbaz} T.,  {Charles} P.~A.,   {Fender} R.~P.,
  2004, \mn@doi [\apjl] {10.1086/425145}, \href
  {http://adsabs.harvard.edu/abs/2004ApJ...613L.133C} {613, L133}

\bibitem[\protect\citeauthoryear{{Casares} et~al.,}{{Casares}
  et~al.}{2009}]{Casares2009}
{Casares} J.,  et~al., 2009, \mn@doi [\apjs] {10.1088/0067-0049/181/1/238},
  \href {http://adsabs.harvard.edu/abs/2009ApJS..181..238C} {181, 238}

\bibitem[\protect\citeauthoryear{{Corral-Santana}, {Casares}, {Shahbaz},
  {Zurita}, {Mart{\'{\i}}nez-Pais}  \& {Rodr{\'{\i}}guez-Gil}}{{Corral-Santana}
  et~al.}{2011}]{Corral-Santana2011}
{Corral-Santana} J.~M.,  {Casares} J.,  {Shahbaz} T.,  {Zurita} C.,
  {Mart{\'{\i}}nez-Pais} I.~G.,   {Rodr{\'{\i}}guez-Gil} P.,  2011, \mn@doi
  [\mnras] {10.1111/j.1745-3933.2011.01022.x}, \href
  {http://adsabs.harvard.edu/abs/2011MNRAS.413L..15C} {413, L15}

\bibitem[\protect\citeauthoryear{{Corral-Santana}, {Casares},
  {Mu{\~n}oz-Darias}, {Rodr{\'{\i}}guez-Gil}, {Shahbaz}, {Torres}, {Zurita}  \&
  {Tyndall}}{{Corral-Santana} et~al.}{2013}]{Corral-Santana2013}
{Corral-Santana} J.~M.,  {Casares} J.,  {Mu{\~n}oz-Darias} T.,
  {Rodr{\'{\i}}guez-Gil} P.,  {Shahbaz} T.,  {Torres} M.~A.~P.,  {Zurita} C.,
  {Tyndall} A.~A.,  2013, \mn@doi [Science] {10.1126/science.1228222}, \href
  {http://adsabs.harvard.edu/abs/2013Sci...339.1048C} {339, 1048}

\bibitem[\protect\citeauthoryear{{Corral-Santana}, {Casares},
  {Mu{\~n}oz-Darias}, {Bauer}  \& {Russell}}{{Corral-Santana}
  et~al.}{2015}]{Corral-Santana2015}
{Corral-Santana} J.~M.,  {Casares} J.,  {Mu{\~n}oz-Darias} T.,  {Bauer} F.~E.,
   {Russell} D.~M.,  2015, submitted to \aap

\bibitem[\protect\citeauthoryear{{Cox}}{{Cox}}{2000}]{Cox2000}
{Cox} A.~N.,  2000, {Allen's astrophysical quantities}

\bibitem[\protect\citeauthoryear{{Dhillon}, {Privett}  \& {Duffey}}{{Dhillon}
  et~al.}{2001}]{Dhillon2001}
{Dhillon} V.~S.,  {Privett} G.~J.,   {Duffey} K.~P.,  2001, Starlink User Note,
  \href {http://adsabs.harvard.edu/abs/2001StaUN.167.....D} {167}

\bibitem[\protect\citeauthoryear{{Falc{\'o}n-Barroso},
  {S{\'a}nchez-Bl{\'a}zquez}, {Vazdekis}, {Ricciardelli}, {Cardiel}, {Cenarro},
  {Gorgas}  \& {Peletier}}{{Falc{\'o}n-Barroso}
  et~al.}{2011}]{Falcon-Barroso2011}
{Falc{\'o}n-Barroso} J.,  {S{\'a}nchez-Bl{\'a}zquez} P.,  {Vazdekis} A.,
  {Ricciardelli} E.,  {Cardiel} N.,  {Cenarro} A.~J.,  {Gorgas} J.,
  {Peletier} R.~F.,  2011, \mn@doi [\aap] {10.1051/0004-6361/201116842}, \href
  {http://adsabs.harvard.edu/abs/2011A%26A...532A..95F} {532, A95}

\bibitem[\protect\citeauthoryear{{Faulkner}, {Flannery}  \&
  {Warner}}{{Faulkner} et~al.}{1972}]{Faulkner1972}
{Faulkner} J.,  {Flannery} B.~P.,   {Warner} B.,  1972, \mn@doi [\apjl]
  {10.1086/180989}, \href {http://adsabs.harvard.edu/abs/1972ApJ...175L..79F}
  {175, L79}

\bibitem[\protect\citeauthoryear{{Fukugita}, {Ichikawa}, {Gunn}, {Doi},
  {Shimasaku}  \& {Schneider}}{{Fukugita} et~al.}{1996}]{Fukugita1996}
{Fukugita} M.,  {Ichikawa} T.,  {Gunn} J.~E.,  {Doi} M.,  {Shimasaku} K.,
  {Schneider} D.~P.,  1996, \mn@doi [\aj] {10.1086/117915}, \href
  {http://adsabs.harvard.edu/abs/1996AJ....111.1748F} {111, 1748}

\bibitem[\protect\citeauthoryear{{Gelino}, {Balman}, {K{\i}z{\i}lo{\v g}lu},
  {Y{\i}lmaz}, {Kalemci}  \& {Tomsick}}{{Gelino} et~al.}{2006}]{Gelino2006}
{Gelino} D.~M.,  {Balman} {\c S}.,  {K{\i}z{\i}lo{\v g}lu} {\"U}.,  {Y{\i}lmaz}
  A.,  {Kalemci} E.,   {Tomsick} J.~A.,  2006, \mn@doi [\apj] {10.1086/500924},
  \href {http://adsabs.harvard.edu/abs/2006ApJ...642..438G} {642, 438}

\bibitem[\protect\citeauthoryear{{Gilmore} \& {Reid}}{{Gilmore} \&
  {Reid}}{1983}]{Gilmore1983}
{Gilmore} G.,  {Reid} N.,  1983, \mnras, \href
  {http://adsabs.harvard.edu/abs/1983MNRAS.202.1025G} {202, 1025}

\bibitem[\protect\citeauthoryear{{Iaria}, {Di Salvo}, {Burderi}  \&
  {Robba}}{{Iaria} et~al.}{2001}]{Iaria2001}
{Iaria} R.,  {Di Salvo} T.,  {Burderi} L.,   {Robba} N.~R.,  2001, \mn@doi
  [\apj] {10.1086/321645}, \href
  {http://adsabs.harvard.edu/abs/2001ApJ...557...24I} {557, 24}

\bibitem[\protect\citeauthoryear{{Jonker} \& {Nelemans}}{{Jonker} \&
  {Nelemans}}{2004}]{Jonker2004}
{Jonker} P.~G.,  {Nelemans} G.,  2004, \mn@doi [\mnras]
  {10.1111/j.1365-2966.2004.08193.x}, \href
  {http://adsabs.harvard.edu/abs/2004MNRAS.354..355J} {354, 355}

\bibitem[\protect\citeauthoryear{{Kallman}, {Angelini}, {Boroson}  \&
  {Cottam}}{{Kallman} et~al.}{2003}]{Kallman2003}
{Kallman} T.~R.,  {Angelini} L.,  {Boroson} B.,   {Cottam} J.,  2003, \mn@doi
  [\apj] {10.1086/345475}, \href
  {http://adsabs.harvard.edu/abs/2003ApJ...583..861K} {583, 861}

\bibitem[\protect\citeauthoryear{{Krimm} et~al.,}{{Krimm}
  et~al.}{2011}]{Krimm2011}
{Krimm} H.~A.,  et~al., 2011, The Astronomer's Telegram, \href
  {http://adsabs.harvard.edu/abs/2011ATel.3138....1K} {3138, 1}

\bibitem[\protect\citeauthoryear{{Kuulkers} et~al.,}{{Kuulkers}
  et~al.}{2013}]{Kuulkers2013}
{Kuulkers} E.,  et~al., 2013, \mn@doi [\aap] {10.1051/0004-6361/201219447},
  \href {http://adsabs.harvard.edu/abs/2013A%26A...552A..32K} {552, A32}

\bibitem[\protect\citeauthoryear{{Marsh}, {Horne}  \& {Shipman}}{{Marsh}
  et~al.}{1987}]{Marsh1987}
{Marsh} T.~R.,  {Horne} K.,   {Shipman} H.~L.,  1987, \mnras, \href
  {http://adsabs.harvard.edu/abs/1987MNRAS.225..551M} {225, 551}

\bibitem[\protect\citeauthoryear{{Mata S{\'a}nchez}, {Mu{\~n}oz-Darias},
  {Casares}, {Steeghs}, {Ramos Almeida}  \& {Acosta Pulido}}{{Mata S{\'a}nchez}
  et~al.}{2015}]{MataSanchez2015}
{Mata S{\'a}nchez} D.,  {Mu{\~n}oz-Darias} T.,  {Casares} J.,  {Steeghs} D.,
  {Ramos Almeida} C.,   {Acosta Pulido} J.~A.,  2015, \mn@doi [\mnras]
  {10.1093/mnrasl/slv002}, \href
  {http://adsabs.harvard.edu/abs/2015MNRAS.449L...1M} {449, L1}

\bibitem[\protect\citeauthoryear{{Mu{\~n}oz-Darias}, {Fender}, {Motta}  \&
  {Belloni}}{{Mu{\~n}oz-Darias} et~al.}{2014}]{Munoz-Darias2014}
{Mu{\~n}oz-Darias} T.,  {Fender} R.~P.,  {Motta} S.~E.,   {Belloni} T.~M.,
  2014, \mn@doi [\mnras] {10.1093/mnras/stu1334}, \href
  {http://adsabs.harvard.edu/abs/2014MNRAS.443.3270M} {443, 3270}

\bibitem[\protect\citeauthoryear{{Orosz} \& {Bailyn}}{{Orosz} \&
  {Bailyn}}{1995}]{Orosz1995}
{Orosz} J.~A.,  {Bailyn} C.~D.,  1995, \mn@doi [\apjl] {10.1086/187930}, \href
  {http://adsabs.harvard.edu/abs/1995ApJ...446L..59O} {446, L59}

\bibitem[\protect\citeauthoryear{{Orosz}, {Bailyn}, {Remillard}, {McClintock}
  \& {Foltz}}{{Orosz} et~al.}{1994}]{Orosz1994}
{Orosz} J.~A.,  {Bailyn} C.~D.,  {Remillard} R.~A.,  {McClintock} J.~E.,
  {Foltz} C.~B.,  1994, \mn@doi [\apj] {10.1086/174962}, \href
  {http://adsabs.harvard.edu/abs/1994ApJ...436..848O} {436, 848}

\bibitem[\protect\citeauthoryear{{Orosz}, {McClintock}, {Aufdenberg},
  {Remillard}, {Reid}, {Narayan}  \& {Gou}}{{Orosz} et~al.}{2011}]{Orosz2011}
{Orosz} J.~A.,  {McClintock} J.~E.,  {Aufdenberg} J.~P.,  {Remillard} R.~A.,
  {Reid} M.~J.,  {Narayan} R.,   {Gou} L.,  2011, \mn@doi [\apj]
  {10.1088/0004-637X/742/2/84}, \href
  {http://adsabs.harvard.edu/abs/2011ApJ...742...84O} {742, 84}

\bibitem[\protect\citeauthoryear{{Ponti}, {Mu{\~n}oz-Darias}  \&
  {Fender}}{{Ponti} et~al.}{2014}]{Ponti2014}
{Ponti} G.,  {Mu{\~n}oz-Darias} T.,   {Fender} R.~P.,  2014, \mn@doi [\mnras]
  {10.1093/mnras/stu1742}, \href
  {http://adsabs.harvard.edu/abs/2014MNRAS.444.1829P} {444, 1829}

\bibitem[\protect\citeauthoryear{{Ponti} et~al.,}{{Ponti}
  et~al.}{2015}]{Ponti2015}
{Ponti} G.,  et~al., 2015, \mn@doi [\mnras] {10.1093/mnras/stu1853}, \href
  {http://adsabs.harvard.edu/abs/2015MNRAS.446.1536P} {446, 1536}

\bibitem[\protect\citeauthoryear{{Rau}, {Greiner}  \& {Filgas}}{{Rau}
  et~al.}{2011}]{Rau2011}
{Rau} A.,  {Greiner} J.,   {Filgas} R.,  2011, The Astronomer's Telegram, \href
  {http://adsabs.harvard.edu/abs/2011ATel.3140....1R} {3140, 1}

\bibitem[\protect\citeauthoryear{{Rayne} \& {Whelan}}{{Rayne} \&
  {Whelan}}{1981}]{Rayne1981}
{Rayne} M.~W.,  {Whelan} J.~A.~J.,  1981, \mnras, \href
  {http://adsabs.harvard.edu/abs/1981MNRAS.196...73R} {196, 73}

\bibitem[\protect\citeauthoryear{{Remillard}, {Orosz}, {McClintock}  \&
  {Bailyn}}{{Remillard} et~al.}{1996}]{Remillard1996}
{Remillard} R.~A.,  {Orosz} J.~A.,  {McClintock} J.~E.,   {Bailyn} C.~D.,
  1996, \mn@doi [\apj] {10.1086/176885}, \href
  {http://adsabs.harvard.edu/abs/1996ApJ...459..226R} {459, 226}

\bibitem[\protect\citeauthoryear{{S{\'a}nchez-Bl{\'a}zquez}
  et~al.,}{{S{\'a}nchez-Bl{\'a}zquez} et~al.}{2006}]{Sanchez-Blazquez2006}
{S{\'a}nchez-Bl{\'a}zquez} P.,  et~al., 2006, \mn@doi [\mnras]
  {10.1111/j.1365-2966.2006.10699.x}, \href
  {http://adsabs.harvard.edu/abs/2006MNRAS.371..703S} {371, 703}

\bibitem[\protect\citeauthoryear{{Schoembs} \& {Hartmann}}{{Schoembs} \&
  {Hartmann}}{1983}]{Schoembs1983}
{Schoembs} R.,  {Hartmann} K.,  1983, \aap, \href
  {http://adsabs.harvard.edu/abs/1983A%26A...128...37S} {128, 37}

\bibitem[\protect\citeauthoryear{{Shafter}, {Szkody}  \&
  {Thorstensen}}{{Shafter} et~al.}{1986}]{Shafter1986}
{Shafter} A.~W.,  {Szkody} P.,   {Thorstensen} J.~R.,  1986, \mn@doi [\apj]
  {10.1086/164549}, \href {http://adsabs.harvard.edu/abs/1986ApJ...308..765S}
  {308, 765}

\bibitem[\protect\citeauthoryear{{Shahbaz}, {Russell}, {Zurita}, {Casares},
  {Corral-Santana}, {Dhillon}  \& {Marsh}}{{Shahbaz}
  et~al.}{2013}]{Shahbaz2013}
{Shahbaz} T.,  {Russell} D.~M.,  {Zurita} C.,  {Casares} J.,  {Corral-Santana}
  J.~M.,  {Dhillon} V.~S.,   {Marsh} T.~R.,  2013, \mn@doi [\mnras]
  {10.1093/mnras/stt1212}, \href
  {http://adsabs.harvard.edu/abs/2013MNRAS.434.2696S} {434, 2696}

\bibitem[\protect\citeauthoryear{{Smith} \& {Dhillon}}{{Smith} \&
  {Dhillon}}{1998}]{SmithDhillon1998}
{Smith} D.~A.,  {Dhillon} V.~S.,  1998, \mn@doi [\mnras]
  {10.1046/j.1365-8711.1998.02065.x}, \href
  {http://adsabs.harvard.edu/abs/1998MNRAS.301..767S} {301, 767}

\bibitem[\protect\citeauthoryear{{Smith}, {Mehes}, {Vande Putte}  \&
  {Hawkins}}{{Smith} et~al.}{2005}]{Smith2005}
{Smith} R.~C.,  {Mehes} O.,  {Vande Putte} D.,   {Hawkins} N.~A.,  2005,
  \mn@doi [\mnras] {10.1111/j.1365-2966.2005.09040.x}, \href
  {http://adsabs.harvard.edu/abs/2005MNRAS.360..364S} {360, 364}

\bibitem[\protect\citeauthoryear{{Steeghs}, {McClintock}, {Parsons}, {Reid},
  {Littlefair}  \& {Dhillon}}{{Steeghs} et~al.}{2013}]{Steeghs2013}
{Steeghs} D.,  {McClintock} J.~E.,  {Parsons} S.~G.,  {Reid} M.~J.,
  {Littlefair} S.,   {Dhillon} V.~S.,  2013, \mn@doi [\apj]
  {10.1088/0004-637X/768/2/185}, \href
  {http://adsabs.harvard.edu/abs/2013ApJ...768..185S} {768, 185}

\bibitem[\protect\citeauthoryear{{Torres} et~al.,}{{Torres}
  et~al.}{2002}]{Torres2002}
{Torres} M.~A.~P.,  et~al., 2002, \mn@doi [\apj] {10.1086/339282}, \href
  {http://adsabs.harvard.edu/abs/2002ApJ...569..423T} {569, 423}

\bibitem[\protect\citeauthoryear{{Torres}, {Callanan}, {Garcia}, {Zhao},
  {Laycock}  \& {Kong}}{{Torres} et~al.}{2004}]{Torres2004}
{Torres} M.~A.~P.,  {Callanan} P.~J.,  {Garcia} M.~R.,  {Zhao} P.,  {Laycock}
  S.,   {Kong} A.~K.~H.,  2004, \mn@doi [\apj] {10.1086/422740}, \href
  {http://adsabs.harvard.edu/abs/2004ApJ...612.1026T} {612, 1026}

\bibitem[\protect\citeauthoryear{{Torres}, {Jonker}, {Miller-Jones}, {Steeghs},
  {Repetto}  \& {Wu}}{{Torres} et~al.}{2015}]{Torres2015}
{Torres} M.~A.~P.,  {Jonker} P.~G.,  {Miller-Jones} J.~C.~A.,  {Steeghs} D.,
  {Repetto} S.,   {Wu} J.,  2015, \mn@doi [\mnras] {10.1093/mnras/stv720},
  \href {http://adsabs.harvard.edu/abs/2015MNRAS.450.4292T} {450, 4292}

\bibitem[\protect\citeauthoryear{{Uemura} et~al.,}{{Uemura}
  et~al.}{2000}]{Uemura2000}
{Uemura} M.,  et~al., 2000, \mn@doi [\pasj] {10.1093/pasj/52.4.L15}, \href
  {http://adsabs.harvard.edu/abs/2000PASJ...52L..15U} {52, L15}

\bibitem[\protect\citeauthoryear{{Vande Putte}, {Smith}, {Hawkins}  \&
  {Martin}}{{Vande Putte} et~al.}{2003}]{Vande2003}
{Vande Putte} D.,  {Smith} R.~C.,  {Hawkins} N.~A.,   {Martin} J.~S.,  2003,
  \mn@doi [\mnras] {10.1046/j.1365-8711.2003.06524.x}, \href
  {http://adsabs.harvard.edu/abs/2003MNRAS.342..151V} {342, 151}

\bibitem[\protect\citeauthoryear{{Zurita}, {Durant}, {Torres}, {Shahbaz},
  {Casares}  \& {Steeghs}}{{Zurita} et~al.}{2008}]{Zurita2008}
{Zurita} C.,  {Durant} M.,  {Torres} M.~A.~P.,  {Shahbaz} T.,  {Casares} J.,
  {Steeghs} D.,  2008, \mn@doi [\apj] {10.1086/588721}, \href
  {http://adsabs.harvard.edu/abs/2008ApJ...681.1458Z} {681, 1458}

\makeatother
\end{thebibliography}


\bsp	
\label{lastpage}
\end{document}